\documentclass[12pt]{article}
\usepackage{setspace}
\usepackage{amsmath}
\usepackage{amssymb}
\usepackage[top=1in,bottom=1in,right=0.5in,left=0.5in]{geometry}
\usepackage{caption}
\usepackage{psfrag}
\usepackage{graphicx}
\usepackage{hyperref}
\usepackage{subfigure}
\usepackage{authblk}
\usepackage{verbatim}
\usepackage{float}
\usepackage{bibcheck}
\usepackage[backend=biber,
            style=numeric,
            sorting=none,
            giveninits=true,
            maxnames=5,
            doi=false,
            isbn=false,
            url=false]{biblatex}
\title{Bouncing shellworld embedded in charged AdS spacetime}
	
\author{Karma P. Sherpa\thanks{E-Mail: sherpa.karma.pincho@gmail.com and karma\_202310028@smit.smu.edu.in} }
\author{Rishi Pokhrel\thanks{E-Mail: rishipokhrel.smit@gmail.com and rishi\_20211037@smit.smu.edu.in} }     
\author{Indra K. P. Chettri\thanks{E-mail: indrapandey.smit@gmail.com and indra\_202410054@smit.smu.edu.in}}                
\author{Tanay K. Dey\thanks{E-mail: tanay.dey@gmail.com and tanay.d@smit.smu.edu.in}}
\affil{Department of Physics, Sikkim Manipal Institute of Technology, Sikkim Manipal University, Majitar, Sikkim-737136, India.}

	\date{}
	
\begin{document}
		\maketitle
		
\begin{abstract}
We investigate the cosmological evolution of a spherical brane within the shellworld (dark bubble) scenario embedded in a five-dimensional charged AdS bulk spacetime. For zero or small value of the brane cosmological constant, the shellworld universe has a nonsingular bouncing and cyclic nature with the bounce induced by the charge parameter of the bulk spacetime. On large value of the brane cosmological constant, the shellworld undergoes a nonsingular bounce and expands eternally. Under certain conditions, the brane bounces outside the bulk horizons, leading to a geometrical resolution of the Cauchy horizon instability present in standard braneworld models. Furthermore, analysis of linear scalar perturbations reveals that test scalar field modes remain regular and non-divergent across the bounce. Finally, we incorporate uniformly distributed long strings in the bulk and demonstrate that the bounce still persists while avoiding the horizon instability.

\end{abstract}

\clearpage

\section{Introduction}
Braneworld models \cite{Randall:1999ee,Randall:1999vf} have provided a new perspective on the nature of our universe by postulating the existence of a higher-dimensional bulk spacetime in which our four-dimensional universe is embedded as a brane. The bulk spacetime is typically taken to be a five-dimensional Anti-de Sitter (AdS) spacetime \cite{Chamblin:1999ya,Kraus:1999it,Nojiri:2002hz}. When the bulk metric is taken to be that of a Reissner-Nordstr\"om AdS black hole, the brane universe exhibits a non-singular bouncing cosmology \cite{Mukherji:2002ft}. The charge parameter in the bulk induces a negative energy density contribution on the brane, which leads to a bounce in the brane universe. However, the brane bounces inside the Cauchy horizon of the charged AdS black hole, which results in an unstable brane universe due to the well-known Cauchy horizon instability \cite{Hovdebo:2003ug}. Several attempts have been made to resolve the Cauchy horizon instability in the braneworld scenario. A brane embedded in a bulk with SU(2) Yang-Mills field led to a bulk without a Cauchy horizon, thus providing a stable bouncing braneworld \cite{Okuyama:2004in}. A stable bouncing braneworld  is achieved via a negative dark radiation term arising in a bulk containing a dilaton and a Kalb–Ramond two-form field \cite{DeRisi:2007dn}. 

A bounce in the braneworld scenario was also achieved by considering a non-charged (AdS Schwarzschild) bulk in the presence of cloud of strings \cite{Sherpa:2026bzw}. The bounce was induced by the negative energy density contribution from the mass parameter of the black hole due to the cloud of strings in the bulk. A negative mass black hole yielded a Cauchy (inner) horizon, and the brane bounced inside the Cauchy horizon, leading to an unstable brane universe. However, using the shellworld model, a bouncing scenario was achieved outside the horizons with a positive mass parameter of the bulk spacetime in presence of a cloud of strings.  The shellworld or the dark bubble model was proposed as an alternative approach to find de-Sitter (dS) space and dark energy in string theory \cite{Banerjee:2018qey}. The model arises from the vacuum transition between two AdS phases, wherein a false AdS vacuum decays to a true AdS vacuum through the nucleation of a spherical bubble wall (brane). Our four-dimensional Universe is embedded on the surface of a spherical shellworld undergoing expansion in the five-dimensional AdS bulk spacetime. As such, the shellworld possesses an inside and an outside region corresponding to the true and false AdS vacua respectively while lacking the usual $\mathbb{Z}_2$ symmetry present in the standard braneworld scenario. The dynamics of the shellworld in the bulk is perceived as a dS universe by an observer on the brane. The shellworld scenario has been widely explored in the literature \cite{Banerjee:2019fzz,Banerjee:2020wix,Banerjee:2020wov,Banerjee:2022ree,Banerjee:2023uto}.

In this work, we consider a shellworld scenario for a charged AdS bulk with and without the presence of string cloud. We first investigate the effects of presence of charge without a string cloud in the bulk on the cosmological evolution of the shellworld universe. The charge of the bulk spacetime induces a negative energy density on the Friedmann equation which leads to a non-singular bouncing and cyclic universe with the scale factor oscillating between a minimum and maximum value in the absence of the brane cosmological constant. Once the brane cosmological constant turns on, there is an another bounce induced by the spatial curvature of the shellworld  at larger value of scale factor. Above a critical value of the brane cosmological constant, the bounce due to the spatial curvature disappears and only the bounce due to the bulk charge parameter exists. However, the nature of the bounce is changed from a cyclic to a single bounce with eternal expansion. Under certain range of the bulk parameters, the bounce occurs outside the horizons of the bulk spacetime, which geometrically resolves the Cauchy horizon instability present in the standard braneworld scenario. Furthermore, the analysis of the bouncing solution under linear test scalar field modes indicates that no divergent amplitudes emerge during the transition through the bounce point. 

We further extend our analysis by introducing a string cloud in the charged AdS bulk spacetime. The string cloud is a collection of one-dimensional objects uniformly distributed in the bulk spacetime \cite{Letelier:1979ej}. The string endpoints are anchored to the brane, while their body is suspended in the radial direction of the bulk spacetime. The string cloud has been studied in various contexts in the literature \cite{Chakrabortty:2011sp,Dey:2017xty,Pokhrel:2023plp,Park:2020jio}. The presence of a string cloud in the bulk spacetime leads to an effective matter contribution on the Friedmann equations of the brane. The bouncing nonsingular nature is still preserved in this case, with the bulk charge inducing the bounce. We also show that the Cauchy horizon instability is geometrically resolved even with the incorporation of a cloud of strings in the bulk spacetime for the shellworld scenario.

The manuscript is organised as follows. We present a concise overview of the shellworld scenario in section \ref{sec2}. Section \ref{sec3} is devoted to the analysis of a bouncing shellworld universe embedded in a charged AdS bulk spacetime, where we demonstrate that the issue of Cauchy horizon instability can be geometrically resolved. In section \ref{sec4}, we extend our analysis by including a cloud of strings in the bulk spacetime that induces a matter contribution on the brane. Lastly, we outline our findings in section \ref{sec5}.

\section{The shellworld scenario}\label{sec2}

The shellworld shares almost similar construction as the Randall-Sundrum (RS) braneworlds except it lacks the $\mathbb{Z}_2$ symmetry of the braneworld model. In the shellworld scenario, the decay of a false AdS vacuum to a true AdS vacuum, through the nucleation of a spherical brane, gives rise to an inside–outside configuration of the bulk spacetime. The dynamics of the brane in the bulk is interpreted as the cosmological evolution of a dS universe by an observer on the brane.

The metric of the five dimensional AdS bulk spacetime on either side of the brane is defined by \cite{Banerjee:2019fzz},
\begin{equation}\label{bulk_metric}
	ds^2_{\pm}=-f_{\pm}(r)dt^2+\frac{dr^2}{f_{\pm}(r)}+r^2 d\Omega_3^2,
\end{equation}
where, 
\begin{equation}\label{Ads_metric}
f_{\pm}(r) = 1 + r^2/L_{\pm}^2,
\end{equation}
represents the spacetime metric of outside and inside region of the spherical brane, $L_\pm$ is the AdS curvature radius with the condition $L_+ > L_-$.
The parameterization of the brane trajectory is given by $r=a(\tau)$ and $t=t(\tau)$, where, $\tau$ is the proper time on the brane. The induced metric on the brane is computed as,
\begin{equation}
    ds^2_{ind}=-d\tau^2+a(\tau)^2 d\Omega_3^2,
\end{equation}
where, $a(\tau)$ is the position in the radial direction of the bulk and scale factor of the shellworld universe. The junction conditions for the shellworld are obtained using the Israel's junction condition \cite{Israel:1966rt},
\begin{equation}
	[K_{ab}]-[K] h_{ab}=-\frac{1}{8\pi G_5} S_{ab}.
\end{equation}
Here, $K_{ab}$ is the extrinsic curvature, $h_{ab}$ denotes the induced metric on the brane, $S_{ab}$ is the brane's stress-energy tensor and $G_5$ is the five dimensional gravitational constant. The absence of $\mathbb{Z}_2$ symmetry reduces the junction conditions for the shellworld with a constant brane tension $\sigma$ to,
\begin{equation}\label{bubble_JC}
\sqrt{f_-(a)+\dot{a}^2}-\sqrt{f_+(a)+\dot{a}^2}=\frac{8 \pi G_5}{3}\sigma a.
\end{equation}
Using (\ref{Ads_metric}) in the above equation, we obtain
\begin{equation}
\sqrt{\frac{1+\dot{a}^2}{a^2}+\frac{1}{L_-^2}}-\sqrt{\frac{1+\dot{a}^2}{a^2}+\frac{1}{L_+^2}}=\frac{8 \pi G_5}{3}\sigma .
\end{equation}
 When four-dimensional effects are weak compared to the underlying five-dimensional scales ($1/a\ll1/L, \dot{a}/a\ll1/L$), the brane approximates a spatially flat Minkowski geometry with the critical brane tension $\sigma_c$,
 \begin{equation}
 \sigma_c=\frac{3}{8\pi G_5}\left(\frac{L_+-L_-}{L_+L_-} \right).
 \end{equation}
 With this brief overview of the shellworld scenario, we now analyse the cosmological evolution of the shellworld embedded in a charged AdS bulk spacetime in the next section.
\section{Charged AdS bulk}\label{sec3}
\begin{figure}[h]
\centering
\includegraphics[width=0.5\linewidth]{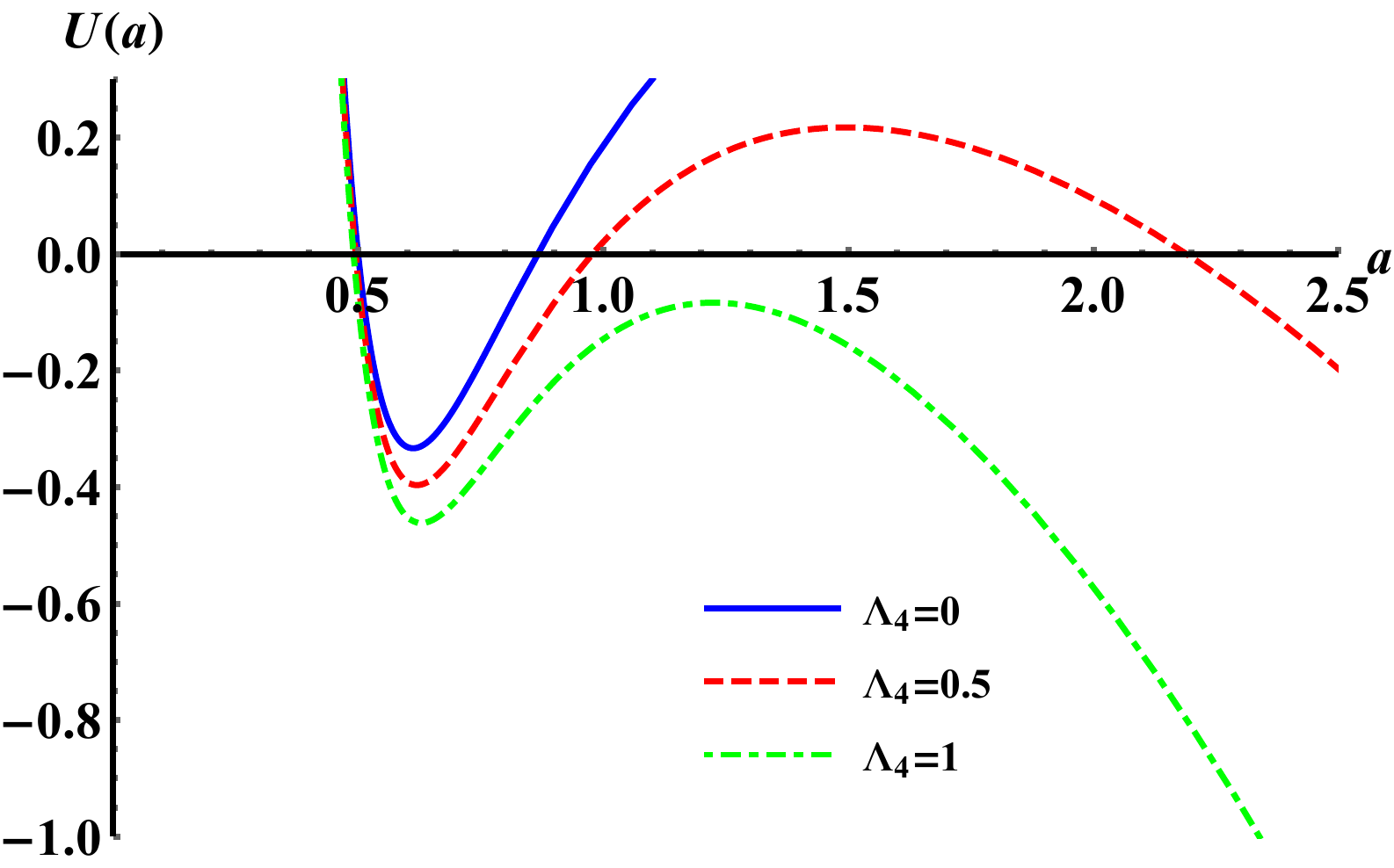}
\caption{Plot of $U(a)$ against $a$ for distinct values of the brane cosmological constant $\Lambda_4$ with the values of the associated parameters as $M_+=1$, $M_-=1$, $L_+=2$, $L_-=1$, $Q_+=1$, $Q_-=1$ and $8\pi G_4=1$.}
\label{fig 1}
\end{figure}

In this section we study the cosmological evolution of the shellworld embedded in the charged AdS black hole spacetime in five dimensions. For this setup, the modified metric function $f_{\pm}(r)$ of equation (\ref{bulk_metric}) for the inside and outside regions of the shellworld can be written from the metric solution of the Reissner-Nordstr\"om AdS black hole \cite{Chamblin:1999tk} and it is expressed as, 
\begin{equation}\label{fr_charge}
f_{\pm}(r)=1+\frac{r^2}{L_{\pm}^2}-\frac{\omega_4 M_{\pm}}{r^2}+\frac{3\omega^2_4 Q_{\pm}^2}{16 r^4},
\end{equation}
where, the parameters $M_\pm$ and $Q_\pm$ are the Arnowitt-Deser-Misner (ADM) mass and charge respectively of the outside and inside region of the spherical brane with the conditions $M_+ > M_-$ and $Q_+ > Q_-$. Further, $\omega_4$ is connected with the three dimensional volume $\omega_3$ via the relation $\omega_4=16 \pi G_5/3 \omega_3$. Using equation (\ref{fr_charge}) in (\ref{bubble_JC}), we obtain the Friedmann equations on the brane as,
\begin{equation}\label{FE_Q}
	H^2=\frac{\dot{a}^2}{a^2}=-\frac{1}{a^2}+\frac{\omega_4}{a^4}\left(\frac{M_+L_+-M_-L_-}{L_+-L_-}\right)-\frac{3\omega_4^2}{16a^6}\left(\frac{Q_+^2L_+-Q_-^2L_-}{L_+-L_-}\right)+\frac{8\pi G_4 }{3}\Lambda_4.
\end{equation}
Here, $H$ is the Hubble factor. The mass and the charge respectively induces an effective dark radiation contribution ($\sim a^{-4}$) and an effective stiff matter contribution ($\sim a^{-6}$) to the Friedmann equation.  $G_4$ and $\Lambda_4$ are the effective four dimensional gravitational and cosmological constant on the brane respectively, given by
\begin{equation}
G_4=\frac{2}{L_+-L_-}G_5,\hspace{0.5cm}\text{and} \hspace{0.5cm} \Lambda_4=\sigma_{c}-\sigma.
\end{equation}
 Using the conformal time $\eta$, defined as $d\tau=a(\eta)d\eta $, the solution of the Friedmann equation (\ref{FE_Q}) for $\Lambda_4=0$ is obtained as,

\begin{equation}\label{scale_factor}
a(\eta)=\sqrt{\frac{\omega_4}{2}\left(\frac{M_+L_+-M_-L_-}{L_+-L_-}\right)}\left[1-\sqrt{1-\left(\frac{3}{4}\frac{(Q_+^2L_+-Q_-^2L_-)(L_+-L_-)}{(M_+L_+-M_-L_- )^2}\right)}\cos2\eta\right]^{\frac{1}{2}}.
\end{equation}
The solution represents a non-singular bouncing universe with the scale factor oscillating between a minimum and maximum value given by,
\begin{equation}
a_{max,min}=\sqrt{\frac{\omega_4}{2}\left(\frac{M_+L_+-M_-L_-}{L_+-L_-}\right)}\left[1\pm\sqrt{1-\left(\frac{3}{4}\frac{(Q_+^2L_+-Q_-^2L_-)(L_+-L_-)}{(M_+L_+-M_-L_- )^2}\right)}\right]^{\frac{1}{2}}.
\end{equation}
The minimal radius of the bounce is obtained at $\eta=n\pi$, where $n$ is an integer. 
The following condition must be satisfied to achieve a non-singular bounce with the incorporation of charge in the bulk,
\begin{equation}\label{con_1}
\left(\frac{3}{4}\frac{(Q_+^2L_+-Q_-^2L_-)(L_+-L_-)}{(M_+L_+-M_-L_- )^2}\right)\leq 1.
\end{equation}
In order to study the nature of the cosmological evolution of the shellworld in the case of $\Lambda_4\neq0$, we analyze the Friedmann equation (\ref{FE_Q}) qualitatively by treating the evolution equation as the dynamics of a point particle moving in an effective potential $U(a)$. From equation (\ref{FE_Q}), the equation of motion can be written as,
\begin{equation}
\dot{a}^2+U(a)=0,
\end{equation}
where, the effective potential $U(a)$ is given as,
\begin{equation}\label{Ua_q}
	U(a)=1-\frac{\omega_4}{a^2}\left(\frac{M_+L_+-M_-L_-}{L_+-L_-}\right)+\frac{3\omega_4^2}{16a^4}\left(\frac{Q_+^2L_+-Q_-^2L_-}{L_+-L_-}\right)-\frac{8\pi G_4 }{3}\Lambda_4 a^2.
\end{equation}
The region $U(a)\leq0$ is the only physically viable region for the cosmological evolution of the shellworld universe while the points where $U(a)=0$ represents the turning points of the scale factor. 
To analyze the evolution of the shellworld universe, we plot equation (\ref{Ua_q}) against the scale factor and observe in figure (\ref{fig 1}) that for the zero value of cosmological constant the universe is bounded within a region which corresponds to a non-singular bouncing and cyclic nature of the universe. Up to a certain range of $\Lambda_4$, there are two regions where the shellworld universe can exist. The first region with smaller $a$ corresponds to a bouncing and cyclic universe induced by the charge parameter while the second region at larger $a$ corresponds to a non-singular expansion induced by the spatial curvature. As the brane cosmological constant increases, the nature of the bounce changes to a single bounce followed by an eternal expansion. \\
However, a non-singular bounce induced by the presence of charge in the AdS bulk spacetime in the braneworld scenario was found to be generally unstable. This is a consequence of the brane bouncing inside the Cauchy horizon of the charged AdS bulk where it encounters a curvature singularity \cite{Hovdebo:2003ug}. Therefore, we now analyse the geometrical stability of the shellworld for a charged AdS bulk spacetime at the bouncing point.
\subsection{Geometric Stability}\label{sec4}
\begin{figure}[h]
	\centering
	\includegraphics[width=0.5\linewidth]{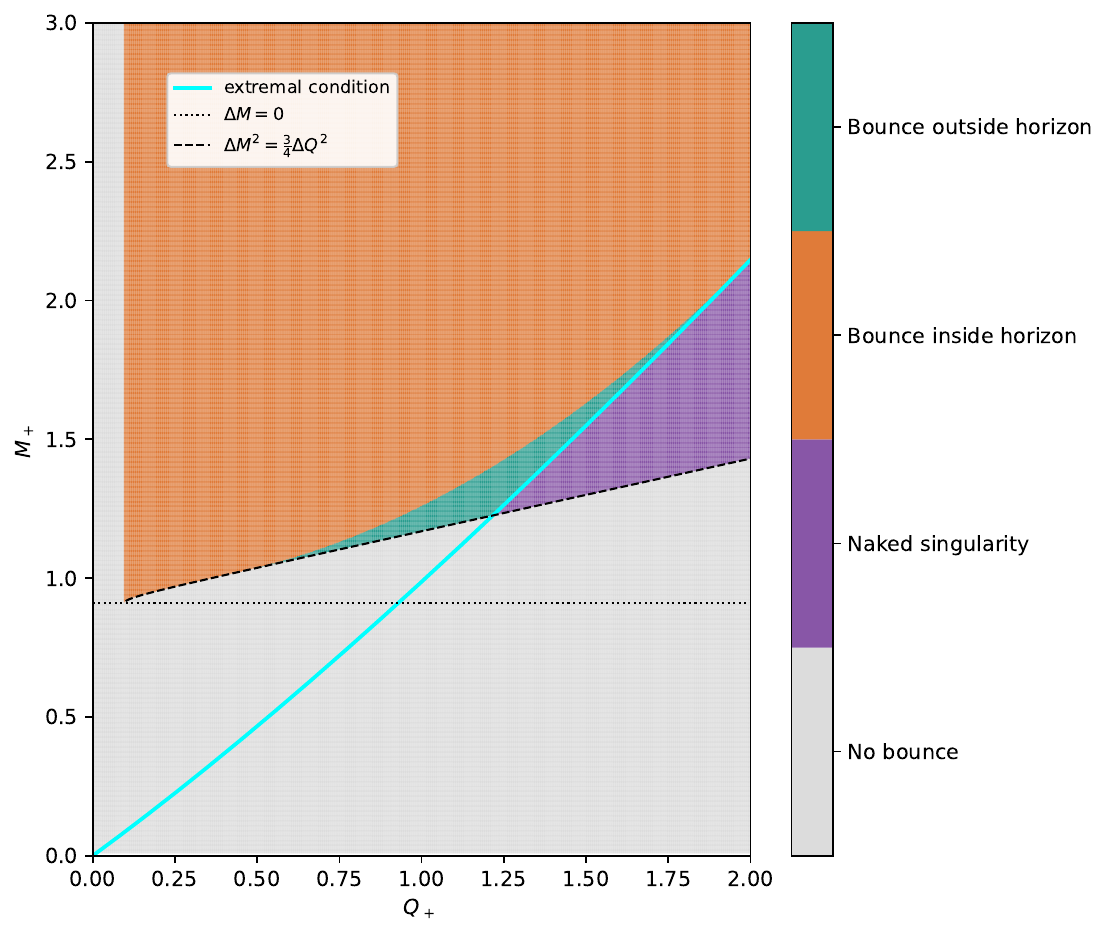}
	\caption{Phase space diagram of the parameters $M_+$ and $Q_+$ for fixed values of the parameters $M_-=1$, $Q_-=0.1$, $L_-=1$ and $L_+=1.1$.}
	\label{fig 2}
\end{figure}

\begin{figure}[h]
	\centering
\subfigure[]{\includegraphics[width=0.45\linewidth]{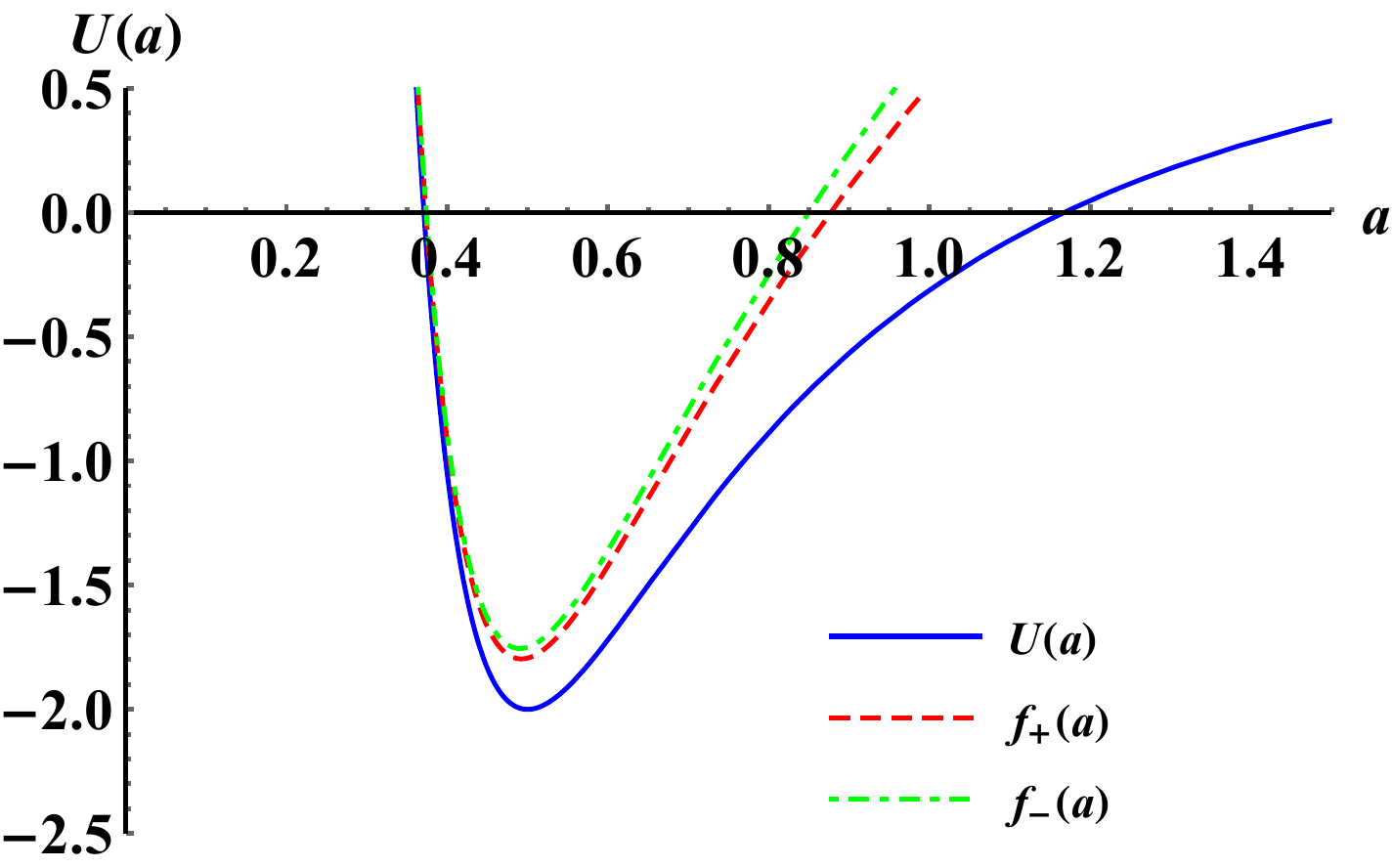}}
\hspace{.2in}\subfigure[]{\includegraphics[width=0.45\linewidth]{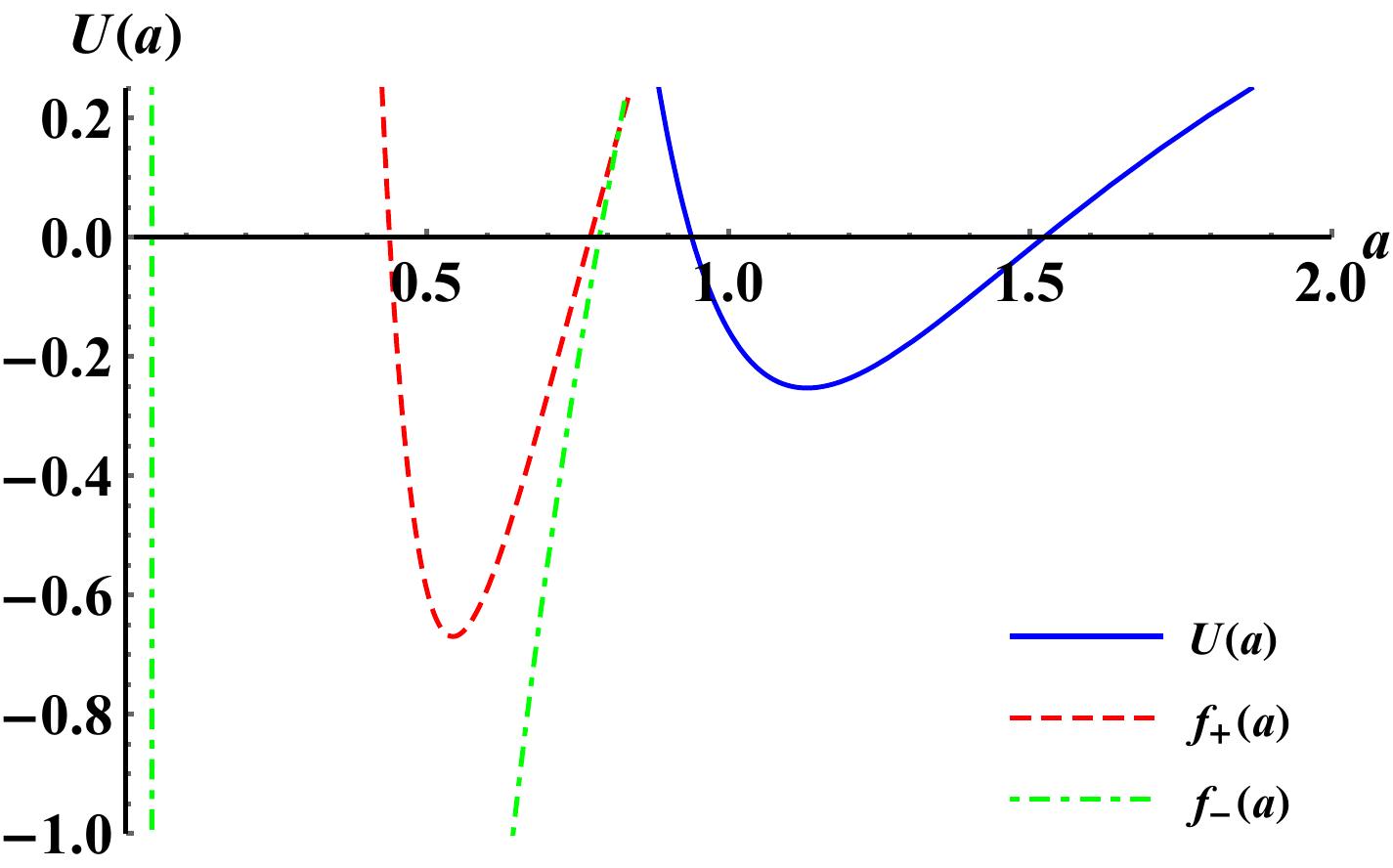}}
	\caption{Plot of $U(a)$ and $f_{\pm}(a)$ against $a$ for distinct values of the parameters. (a) represents bounce inside the Cauchy horizon with the value of the parameters $M_+=M_-=1.5$, $Q_+=Q_-=1$, $L_+=1.1$, $L_-=1$ and $\Lambda_4=0$ while (b) represents bounce outside the horizons with the value of the parameters $M_+=1.2$, $M_-=1$, $Q_+=1$, $Q_-=0.1$, $L_+=1.1$, $L_-=1$ and $\Lambda_4=0$. }
	\label{fig 3}
\end{figure}

To achieve a viable bounce in the shellworld scenario, the bounce should occur outside both the horizons of the bulk spacetimes \cite{DeRisi:2007dn}. The scale factor at the bounce can be obtained by solving equation (\ref{Ua_q}) for $U(a)=0$. Since the bounce induced by the charge parameter is the smallest root as shown in figure (\ref{fig 1}), we can set the effective brane cosmological constant to zero, i.e., $\Lambda_4=0$ to analyse the stability. Thus, the condition for a geometrically stable bounce is given by $a_b>a_{h(o)}$, where, $a_b$ is the scale factor at the bounce induced by the charge parameter and $a_{h(o)}$ is the outer horizon.  The scale factor at the bounce is given by the roots of the following equation,
\begin{equation}
	a_b^4-\omega_4\left(\frac{M_+L_+-M_-L_-}{L_+-L_-}\right)a_b^2+\frac{3\omega_4^2}{16}\left(\frac{Q_+^2L_+-Q_-^2L_-}{L_+-L_-}\right)=0.
\end{equation}
By Descartes' rule of signs, we find the equation to have either two or no positive roots. The smallest positive root of the above equation is
\begin{equation}
	a_b=
\sqrt{\frac{\omega_4}{2}\left[\frac{M_+L_+-M_-L_-}{L_+-L_-}-\sqrt{\left(\frac{M_+L_+-M_-L_-}{L_+-L_-}\right)^2-\frac{3}{4}
\left(\frac{Q_+^2L_+-Q_-^2L_-}{L_+-L_-}\right)
}\right]}.
\end{equation}
The horizons of the bulk spacetime for a charged black hole are obtained by using the condition $f(a)=0$ in (\ref{fr_charge}), which results in
\begin{equation}\label{6_poly}
	\frac{a_h^6}{L_{\pm}^2}+a_h^4-\omega_4 M_{\pm} a_h^2+\frac{3\omega_4^2 Q_{\pm}^2}{16}=0.
\end{equation}
By using Cardano's method for solving the above equation, we find the largest real root as
\begin{equation}
a_{h(o)}=\sqrt{\frac{L_\pm^2}{3}\left(\sqrt[3]{\frac{2}{A}} \left( 1+\frac{3 \omega_4 M_\pm}{L_\pm^2}\right)+\sqrt[3]{\frac{A}{2}}-1\right)}
\end{equation}
where,
\begin{equation}
A=-\left(
\frac{81\omega_4^2Q_\pm^2}{16L_\pm^4}+\frac{9\omega_4 M_\pm}{L_\pm^2}+2\right)+\sqrt{\left(
\frac{81\omega_4^2Q_\pm^2}{16L_\pm^4}
+\frac{9\omega_4 M_\pm}{L_\pm^2}
+2\right)^2-4\left(\frac{3\omega_4 M_\pm}{L_\pm^2}+1\right)^3}.
\end{equation}
The condition for the bounce to occur outside the horizon can be expressed as
\begin{equation}\label{con_2}
\begin{split}
\frac{\omega_4}{2}
\left[
\frac{M_+L_+-M_-L_-}{L_+-L_-}
-\sqrt{
\left(\frac{M_+L_+-M_-L_-}{L_+-L_-}\right)^2
-\frac{3}{4}
\left(\frac{Q_+^2L_+-Q_-^2L_-}{L_+-L_-}\right)
}
\right]
> \\
\frac{L_\pm^2}{3}\left(\sqrt[3]{\frac{2}{A}} \left( 1+\frac{3 \omega_4 M_\pm}{L_\pm^2}\right)+\sqrt[3]{\frac{A}{2}}-1\right).
\end{split}
\end{equation}
Thus, the conditions presented in equations (\ref{con_1}) and (\ref{con_2}) are required to obtain a bounce outside the bulk horizons in the shellworld scenario for a charged AdS bulk. However, from the conditions presented above, it is difficult to find the regions for the bounce to occur outside the horizons. We thus resort to a numerical analysis of the conditions from where we can plot a phase space diagram of the parameters such that, we can determine the values needed to find a suitable bounce. This can be done by fixing the values of the parameters $M_-$, $Q_-$, $L_-$ and $L_+$ and varying the values of $M_+$ and $Q_+$ to find the regions where the bounce occurs outside the horizons. The phase space diagram between $M_+$ and $Q_+$ is shown in figure (\ref{fig 2}) where the green region represents the parameter space for which the bounce occurs outside the horizons of the bulk spacetime. The orange region represents the parameter space for which the bounce occurs inside the horizons of the bulk spacetime. The grey region represents the parameter space for which no bounce occurs while the purple region represents the parameter space where the bulk encounters a naked singularity. The blue line represents the condition of an extremal black hole in the bulk spacetime, i.e., the outer and inner horizons merges into a single horizon. The dotted line indicated as $\Delta M=0$, represents
\begin{equation}
\Delta M=\frac{M_+L_+-M_-L_-}{L_+-L_-}=0,
\end{equation}
while, the dashed line indicated as $\Delta M^2-\frac{3}{4}\Delta Q^2=0$, represents
\begin{equation}
\Delta M^2-\frac{3}{4}\Delta Q^2=\left(\frac{M_+L_+-M_-L_-}{L_+-L_-}\right)^2-\frac{3}{4}\left(\frac{Q_+^2L_+-Q_-^2L_-}{L_+-L_-}\right)=0.
\end{equation}
Through the values of the parameters in the green region of figure (\ref{fig 2}), we can achieve a bounce outside the horizons of the bulk spacetime.
We thus plot the effective potential $U(a)$ along with $f_{\pm}(a)$ against the scale factor to analyze the geometric stability of the bounce.
It is evident from figure (\ref{fig 3}) that under the values of the parameters from the green region, the shellworld can avoid the Cauchy horizon instability in the shellworld scenario as the brane bounces outside both of the horizons of the bulk spacetime. Thus, the bouncing mechanism is found to be free from geometric instabilities. However, the issue of perturbative stability remains an open question in bouncing cosmological models and it is discussed in the next subsection.
\subsection{Evolution of Test Scalar Fluctuations}\label{sec5}
In principle, a complete analysis for the stability of the bouncing cosmological model requires a treatment of perturbations in the full five-dimensional background, however, such an approach to the present setup is highly nontrivial. Nevertheless, it is still possible to investigate the stability of the bouncing solution within an effective approach.

To examine the viability of the bouncing solution, we analyze its behavior under linear scalar perturbations. We study the fluctuations of a massless scalar field $\phi$ with minimal coupling confined to the brane, treating it as a test field whose contribution to the background cosmological evolution is negligible. The Fourier modes of the field fluctuation obey the linearized Klein–Gordon equation. Upon defining the canonical variable $v_k\equiv a\delta \phi_k$ for a mode of fluctuations $\delta \phi_k$, the Mukhanov-Sasaki equation reduces to,
\begin{equation}\label{MS}
	v_k'' + (k^2-\frac{a''}{a})v_k=0.
\end{equation}
Here, $k$ is the comoving momentum and $(')$ represents differentiation with respect to the conformal time $\eta$. The scale factor of equation (\ref{scale_factor}) can be written as,
\begin{equation}
a(\eta)=\xi\left[1-\beta \cos2\eta\right]^{\frac{1}{2}}, 
\end{equation}
where,
\begin{equation}
	\xi=\sqrt{\frac{\omega_4}{2}\left(\frac{M_+L_+-M_-L_-}{L_+-L_-}\right)}\hspace{0.2 cm} \text{and}\hspace{0.2 cm}\beta=\sqrt{1-\left(\frac{3}{4}\frac{(Q_+^2L_+-Q_-^2L_-)(L_+-L_-)}{(M_+L_+-M_-L_- )^2}\right)}.
\end{equation}
For the above scale factor, the term $a''/a$ in equation (\ref{MS}) reduces to,
\begin{equation}\label{}
	\frac{a''}{a}=\frac{2\beta \cos 2 \eta -\beta^2(1+\cos^22\eta)}{(1-\beta \cos 2 \eta)^2}.
\end{equation} 
The minimum of the scale factor or the point of bounce occurs at $\eta=n \pi$ for $\beta<1$. The above expression at the point of bounce reduces to $a''/a=2\beta/(1-\beta)$ and remains finite and doesn't diverge at the bounce. At short wavelength regime $(k^2\gg a''/a)$, solving the Mukhanov-Sasaki equation (\ref{MS}) gives us the fluctuation $\delta \phi_k$ as, 
\begin{equation}
	\delta \phi_k \equiv \frac{v_k}{a}= \frac{ e^{-ik\eta}}{\sqrt{2k}a},			
\end{equation}
which represents oscillatory modes.
 For the long wavelength regime $(k^2\ll a''/a)$, the term $a''/a$ dominates over $k^2$, which results in,
\begin{equation}
	\delta \phi_k\equiv C_1 +C_2 \int \frac{d\eta}{a^2(\eta)}.
\end{equation}
The integral term in the above equation can be evaluated as,
\begin{equation}
	\int \frac{d\eta}{a^2(\eta)}=\frac{1}{\xi^2}\int \frac{d\eta}{1-\beta \cos 2 \eta}=\frac{1}{\xi^2\sqrt{1-\beta^2}}\arctan\left(\sqrt{\frac{1+\beta}{1-\beta}}\tan \eta\right).
\end{equation}
The integral is regular in the vicinity of the bounce and therefore does not introduce any divergence in the perturbation variables.

Since $a(\eta)$ remains non-zero at the bounce and the integral $\int d \eta/a^2(\eta)$ is regular in its vicinity, both $v_k$ and the scalar field fluctuations $\delta \phi_k$ remain finite throughout the evolution. Therefore, no divergence develops in the linear perturbation modes of the test scalar field across the bounce. This indicates that the bouncing background is perturbatively well behaved with respect to scalar-field fluctuations. A complete stability analysis would require studying the full set of coupled scalar and metric perturbations, along with their backreaction on the background spacetime. Such an analysis is highly non-trivial and the detailed study of this aspect is deferred to future work. Nevertheless, the regular behavior of the perturbation modes across the bounce serves as an important consistency check and provides encouraging evidence that the shellworld scenario provides an effective approach for a bouncing cosmological model where the bounce is induced by the charge of the black hole. 
\section{Charged AdS bulk with string cloud}\label{sec4}
\begin{figure}[h]	
	\centering
	\includegraphics[width=0.5\linewidth]{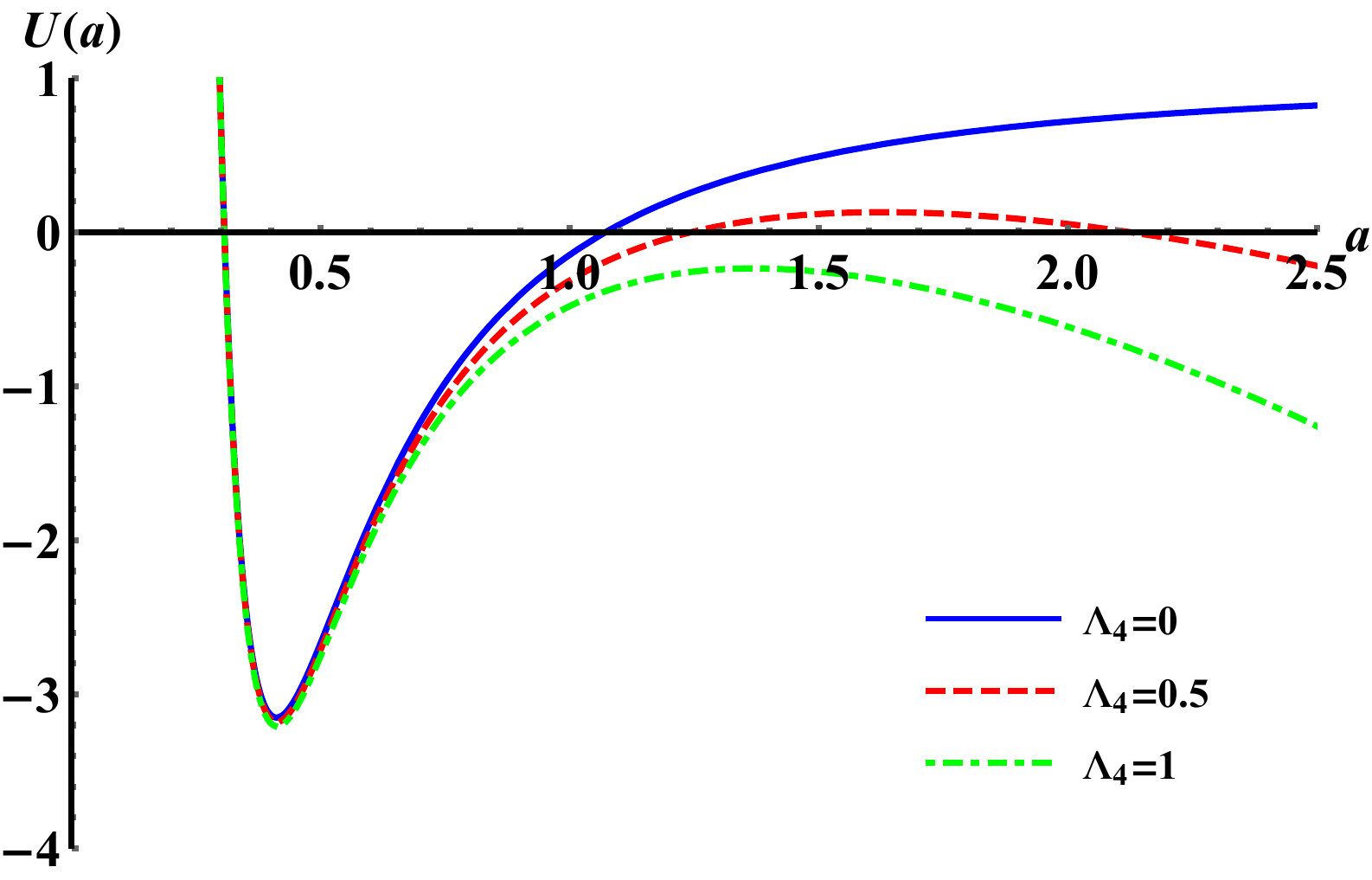}
	\caption{Plot of $U(a)$ against $a$ for distinct values of the brane cosmological constant $\Lambda_4$ with the values of the associated parameters as $M_+=1$, $M_-=1$, $L_+=2$, $L_-=1$, $Q_+=1$, $Q_-=1$, $b_+=1$, $b_-=1$ and $8\pi G_4=1$.}
	\label{fig 4}		
	\end{figure}

We now discuss a matter contribution in the Friedmann equations on the shellworld. The matter is incorporated by the string cloud in the bulk spacetime. The strings are extended in the radial direction of the bulk spacetime and uniformly distributed. End points of the strings are anchored to the spherical shellworld. The string endpoints represents massive particles in the brane universe, and are interpreted as induced matter contribution on the brane. The function $f_{\pm}(r)$ for the inside and outside regions of the shellworld in this case, can be written from the metric solution $f(r)$ of charged black hole with string cloud \cite{Pokhrel:2023plp} and it is expressed as,
\begin{equation}
f_{\pm}(r)=1+\frac{r^2}{L_{\pm}^2}-\frac{\omega_4 M_{\pm}}{r^2}+\frac{3\omega^2_4 Q_{\pm}^2}{16 r^4}-\frac{16\pi G_5 b_{\pm}}{3r}.
\end{equation}
Here, $b$ represents the density of the string cloud.
 Using the above equation in (\ref{bubble_JC}), we obtain the Friedmann equations on the brane as,
\begin{equation}\label{FE_w_string}
\frac{\dot{a}^2}{a^2}=-\frac{1}{a^2}+\frac{\omega_4}{a^4}\left(\frac{M_+L_+-M_-L_-}{L_+-L_-}\right)-\frac{3\omega_4^2}{16a^6}\left(\frac{Q_+^2L_+-Q_-^2L_-}{L_+-L_-}\right)+\frac{8\pi G_4}{3a^3}\left(b_+L_+-b_-L_-\right)+\frac{8\pi G_4 }{3}\Lambda_4.
\end{equation}
This shows that the string cloud contributes an effective matter term ($\sim a^{-3}$) induced by the bulk geometry to the Friedmann equations on the brane. We conduct a qualitative analysis of the Friedmann equation as in the preceding section to study the nature of the cosmological evolution of the shellworld in this case, and the effective potential $U(a)$ appears as,
\begin{equation}\label{Ua_b}
U(a)=1-\frac{\omega_4}{a^2}\left(\frac{M_+L_+-M_-L_-}{L_+-L_-}\right)+\frac{3\omega_4^2}{16a^4}\left(\frac{Q_+^2L_+-Q_-^2L_-}{L_+-L_-}\right)-\frac{8\pi G_4}{3a^3}\left(b_+L_+-b_-L_-\right)-\frac{8\pi G_4 }{3}\Lambda_4 a^2.
\end{equation}
The evolution of the shellworld universe is studied by plotting equation (\ref{Ua_b}) with respect to the scale factor in figure (\ref{fig 4}).
We observe that like the previous case, the bounce persists for distinct values of the effective cosmological constant on the brane even with the incorporation of string cloud in the bulk spacetime. For $\Lambda_4=0$, we obtain a nonsingular bouncing and cyclic universe while for a certain range of $\Lambda_4$, we obtain two regions as in the previous case where a bouncing and cyclic nature is observed for smaller $a$, while a non-singular eternal expansion is observed for larger $a$. For large value of $\Lambda_4$, we obtain a single bounce followed by an eternal expansion.\\

\begin{figure}[h]
	\centering
	\subfigure[]{\includegraphics[width=0.45\linewidth]{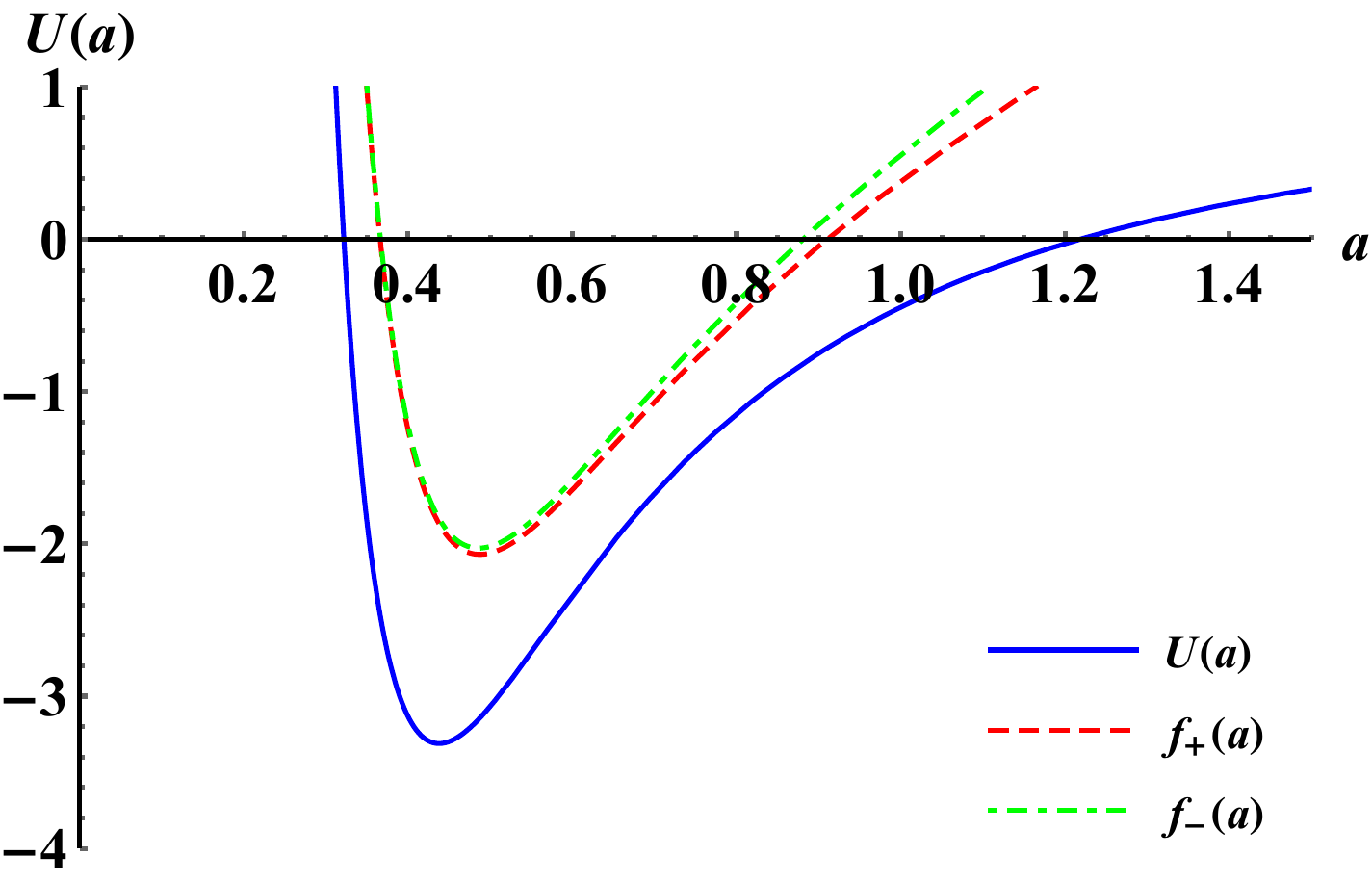}}
	\hspace{.2in}\subfigure[]{\includegraphics[width=0.45\linewidth]{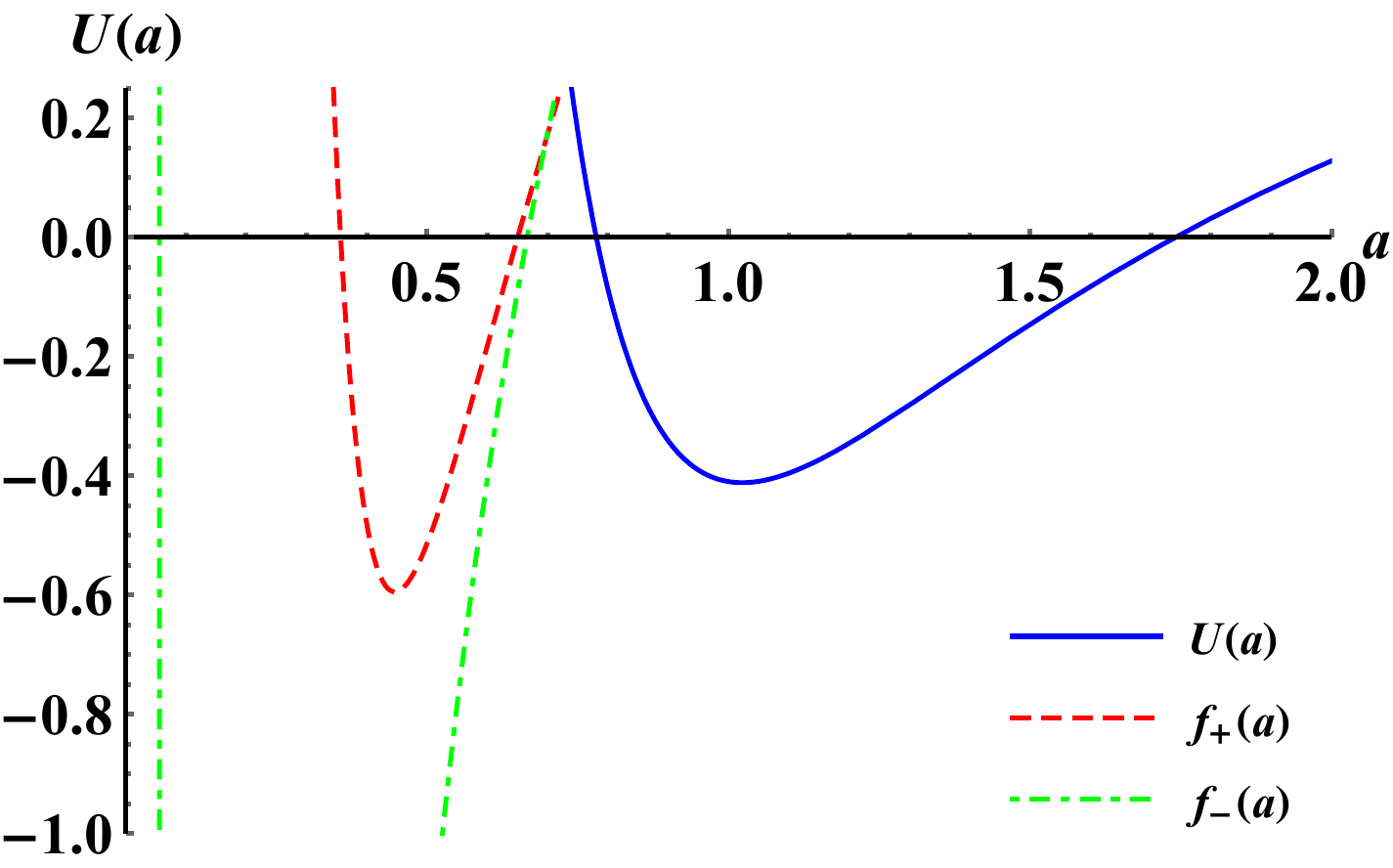}}
	\caption{Plot of $U(a)$ and $f_{\pm}(a)$ against $a$ with the incorporation of string cloud. (a) represents bounce inside the horizons with value of the parameters as $M_+=M_-=1.5$, $Q_+=Q_-=1$, $L_+=1.1$, $L_-=1$, $b_+=b_-=0.2$, $\Lambda_4=0$ and $8\pi G_4=1$; while (b) represents bounce outside the horizons with value of the parameters as $M_+=M_-=0.6$, $Q_+=0.6$, $Q_-=0.1$, $L_+=1.1$, $L_-=1$, $b_+=0.3$, $b_-=0.1$, $\Lambda_4=0$ and $8\pi G_4=1$.}
	\label{fig 5}
\end{figure}
Since obtaining analytical solutions of the polynomial equations corresponding to $f_{\pm}(a)=0$ and $U(a)=0$ is highly nontrivial in this case, we are unable to determine the exact conditions for the bounce to occur outside the horizons as was done in the previous section. We plot the effective potential $U(a)$ along with $f_{\pm}(a)$ against the scale factor to analyze the geometric stability of the bounce for distinct values of the parameters in figure (\ref{fig 5}). From figure \ref{fig 5}, we observe that a bounce outside the horizons of the bulk spacetimes can be achieved even with the incorporation of string cloud in the charged bulk spacetime under certain tuning of the parameters.

\section{Summary}\label{sec5}
In this work, we have studied the cosmological evolution for the shellworld scenario in a charged AdS bulk with/without a presence of string cloud. Initially, we analyse the configuration in the absence of string cloud. We derive the Friedmann equation for the shellworld where its solution represents a non-singular bouncing and cyclic universe with the scale factor oscillating between a minimum and maximum value in the absence of the brane cosmological constant. The bounce is induced by the charge parameter of the bulk spacetime. Once the value of the brane cosmological constant increases, the nature of the bounce changes from a cyclic bounce to a single bounce followed by an eternal expansion. Under certain conditions, the brane bounces outside the horizons of the bulk spacetime, thus providing a geometric resolution to the Cauchy horizon instability present in the standard braneworld scenario. We also examine the behavior of linear scalar perturbations in the bouncing background and our analysis confirms that linear fluctuations of a minimally coupled test scalar field remain regular across the bounce. We have further extended our analysis by including a matter contribution from a uniformly distributed long strings in the bulk spacetime where we find the brane bounces outside the bulk horizons even with the incorporation of string cloud in the bulk spacetime, thus geometrically resolving the Cauchy horizon instability in this case as well. Thus, we conclude that the shellworld model provides an effective approach for bouncing scenarios in the charged AdS bulk spacetime with/without a presence of string cloud.

\section*{Acknowledgements}
K. P. Sherpa would like to thank the Ministry of Tribal Affairs, Government of India for the financial support through NFST. R.P. and I.K.P.C. acknowledges the TMA-PAI research grant provided by Sikkim Manipal University.

\printbibliography

\end{document}